Thermo-magnetic characterization of phase transitions in a Ni-Mn-In metamagnetic shape memory alloy

F. J. Romero, J. M. Martín-Olalla, J. S. Blázquez, M. C. Gallardo,

Departamento de Física de la Materia Condensada, Facultad de Física, ICMSE-CSIC, Universidad de Sevilla, Avenida Reina Mercedes s/n, ES41080 Sevilla, Spain

Daniel Soto Parra,

Tecnológico Nacional de México/Instituto Tecnológico de Delicias, Paseo Tecnológico km. 3.5, Cd. Delicias, Chihuahua 33000, Mexico

Eduard Vives and Antoni Planes

Departament de Física de la Matèria Condensada, Facultat de Física, Martí i Franquès 1, Universitat de Barcelona, 08028 Barcelona, Catalonia

ABSTRACT

The partially overlapped ferroelastic/martensitic and para-ferromagnetic phase transitions of a $Ni_{50.53}Mn_{33.65}In_{15.82}$ metamagnetic shape memory alloy have been studied from calorimetric, magnetic and acoustic emission measurement.

We have taken advantage of the existence of thermal hysteresis of the first order ferroelastic/martensitic phase transition (~2.5K) to discriminate the latent heat contribution $\Delta H_t = 7.21(15)$ kJkg$^{-1}$ and the specific heat contribution $\Delta H_c = 216(1)$ Jkg$^{-1}$ to the total excess enthalpy of the phase transition. The specific heat was found to follow a step-like behavior at this phase transition.

The intermittent dynamics of the ferroelastic/martensitic transition has been characterized as a series of avalaches detected both from acoustic emission and calorimetric measurements. The energy distribution of these avalanche events was found to follow a power law with a characteristic energy exponent $\varepsilon \cong 2$ which is in agreement with the expected value for a system undergoing a symmetry change from cubic to monoclinic.

Finally, the critical behavior of the para-ferromagnetic austenite phase transition that takes place at ~311K has been studied from the behavior of the specific heat. A critical exponent $\alpha \cong 0.09$ has been obtained, which has been shown to be in agreement with previous values reported for Ni-Mn-Ga alloys but different from the critical divergence reported for pure Ni.

1. INTRODUCTION

Ni-Mn-In belongs to the family of magnetic Heusler materials that undergoes a martensitic/ferroelastic transition[1]. This class of materials, in addition to usual mechanical shape-memory and super elastic behavior[2], has the ability to recover from giant deformations (sometimes reaching up to 10%) that can be induced by application of a moderate magnetic field. This can occur by means of the following two mechanisms. First, through the symmetry change taking place at the ferroelastic transition, which requires that the transition can be induced by application of a magnetic field. This is the dominant mechanism in the so-called metamagnetic shape memory materials and is feasible when magnetic and structural degrees of freedom are strongly coupled. Second, through a magnetic field induced rearrangement of martensitic/ferroelastic variants (or domains), that requires that the martensitic phase has a large enough uniaxial magnetocrystalline anisotropy, that favors minimization of Zeeman energy by twin boundary motion instead of rotation of magnetic moments. This occurs in the so-called magnetic shape-memory materials, such as the $Ni_2MnGa$ compound. Ni-Mn-In is the prototypical example of metamagnetic shape memory alloy. Therefore, this material can be defined as magneto-structural multiferroic, which shows cross-response to mechanical and magnetic fields. Besides mechanical and magnetic shape memory properties, it displays many other interesting functional properties, and particularly excellent caloric and multicaloric effects[3], which have been acknowledged as potentially interesting for solid state refrigeration applications.

In a certain range of compositions, away from the 2-1-1 full Heusler composition, Ni-Mn-In shows an $L2_1$ (cubic), paramagnetic structure at high temperature. On cooling, it becomes ferromagnetic at the Curie temperature and, at a lower temperature, undergoes a martensitic transition to a 10M monoclinic phase[4], whose magnetic character is controversial. Although a paramagnetic state for this phase has been considered [4], other authors have identified magnetic blocking or antiferromagnetic states [5–7]. In any case, the specific behavior seems to be very sensitive to specific composition effects. These martensitic transitions are known to occur intermittently through localized strain relaxation events, which often display avalanche criticality[8]. This class of critical behavior manifests itself in a power law distribution of the avalanche energies, characterized by an exponent that depends, to a large extent, on the change of symmetry taking place at the transition [9]. In Heusler alloys this behavior has been reported to occur in $Ni_2MnGa$ [10], but not in Ni-Mn-In. Below the martensitic transition, Ni-Mn-In undergoes a further magnetic transition to a low temperature ferromagnetic phase[4].

Specific features of the martensitic transition are controlled by the magnetostructural interplay, which is originated from a fine energy balance decided by the competition of atom displacements associated with the structural change and the highly oscillating RKKY-type magnetic exchange. The strength of this interplay is more intense as the martensitic transition occurs closer to the Curie point. Therefore, in these materials, functional properties associated with the cross response to mechanical and magnetic field are optimized when both martensitic and high temperature para-ferromagnetic transitions occur almost simultaneously. However, in this situation, due to overlapping effects between both transitions, an accurate characterization of thermo-magnetic properties of these transitions is difficult. The present paper aims at characterizing these transitions from magnetic, calorimetric and acoustic emission measurements in a Ni-Mn-In single crystal where they occur with some overlapping. In particular, this is performed by taking advantage of the existence of hysteresis in the martensitic transition that has enabled us to accurately determine the heat capacity difference between the high temperature ferromagnetic austenite phase and the martensitic phase in its paramagnetic state, and the critical behavior of the high temperature ferromagnetic transition that takes place at the Curie point of the austenite phase.

## 2. EXPERIMENTAL

The sample was prepared from the appropriate quantities of high-purity nickel, manganese, and indium. The metals were arc melted several times under an argon atmosphere, flipping the buttons each time. The buttons were then remelted and the alloy drop cast into a copper chill cast mold to ensure compositional homogeneity throughout the ingots. The crystals were grown in a resistance furnace from the as-cast ingots in an alumina Bridgman style crucible.

An alloy composition of $Ni_{50.53}Mn_{33.65}In_{15.82}$ was revealed from energy dispersive X-ray fluorescence measurements (spectrometer EDAX µFRXEAGLE III) at the Research, Technology and Innovation Centre (CITIUS, University of Sevilla, Spain).

Specific magnetization measurements as a function of the applied magnetic field, $\sigma(H)$, up to $\mu_0 H = 1.5$ T were obtained at different temperatures in a vibrating sample magnetometer (Lakeshore 7407) equipped with a LN2 cryostat. The sample mass (1.533 mg) was measured using a high precision microbalance (±0.001 mg). In the vicinity of the structural transition (in the temperature range from 260 to 352 K), the sample was first cooled down to 150 K at zero field and then heated up to measurement temperature (ZFC measurement protocol). This protocol assures a fully martensite state for each magnetization curve obtained below (but close to) the austenite start temperature at zero field. In fact, after magnetization has induced formation of austenite, hysteretic behavior ascribed to the transition makes that removing the field cannot yield a zero austenite fraction[11]. Complementary magnetization curves were obtained from the sample after measuring at 352 K and cooling from 250 K to 170 K to register the Curie transition of the martensite phase. This latter procedure led to a tiny amount of remnant austenite phase at low temperature.

Measurements of heat flux and specific heat were performed using a high-resolution conduction calorimeter[12]. The sample for calorimetry is nearly octagonal with flat faces of area 90 mm$^2$ and 1.89 mm height. The mass of the sample is 1.5487 g.

Specific heat is measured using the method previously described in Ref. [13]. A constant power is dissipated in both heaters during a time long enough (12 min in present experiments) to reach a steady state characterized by a constant temperature difference between the sample and the calorimeter block (less than 0.05 K). Power is then cut off until a new steady state is reached. The power is then switched on again and the sequence is continuously repeated while the temperature of the assembly is changed at a constant low rate (typically 0.1 K/h). Integration of the electromotive force given by the fluxmeters, proportional to the heat flux, between every pair of steady state distribution provides the thermal capacity of the sample.

The equipment can also work like a very sensitive differential thermal analysis (DTA) device. A DTA trace is continuously measured in a different run without dissipation in the sample. The electromotive force given by the fluxmeters is proportional to the heat flux, the integral of which, in turn, is proportional to the total enthalpy change of the sample[14].

For a complete characterization of avalanche criticality, acoustic emission (AE) measurements have been carried out. The sample was mounted on top of a copper block situated inside a double Faraday cage constituted of cooper and iron shielding layers. It was heated and cooled by circulating a temperature-controlled fluid, which allows to establish linear temperature ramps with selected rates in the range from $10^{-3}$ K/s to $10^{-1}$ K/s. The AE was detected with a piezoelectric transducer (micro-80 transducer, encapsulated in stainless steel) acoustically coupled to the upper surface of the sample by a thin layer of petroleum jelly. The output voltage of the transducer was amplified by 60 dB, band filtered between 100 kHz and 400 kHz and transferred to a PCI-2 digital acquisition system (Europhysical Acoustics). Data were stored at a sampling rate of 10 MHz. For identification of AE events associated with avalanches, a threshold above the instrument noise was fixed. Then AE event, identified with avalanches, are assumed to start when the signal crosses for the first time the voltage threshold at a time $t_i$. The event will end at $t_f = t_i + \Delta t_i$ corresponding to the time at which the voltage crosses the threshold in the downwards direction and remains below the threshold for more than a preselected time (or hit detection time, HDT) that was set to 100 µs in our experiments. The energy of each event was estimated by numerical integration of the square voltage during the duration $\Delta t_i$, normalized by a reference electric resistance, $R$ (=10 kΩ, in our case).

3. RESULTS AND DISCUSSION

Figure 1 shows the dependence of the different transition temperatures in Ni-Mn-In alloys on the electrons per atom $e/a$ taken from different literature sources [4,15,16], $e/a$ is calculated as the concentration-weighted average of the valence electrons, which are the electrons in the external s, p and partially occupied d orbitals of the constituents. The ferromagnetic (FM) Curie temperatures of the austenitic and martensitic states are given as $T_C^A$ and $T_C^M$, respectively, and the martensite start temperature is given as $T_S^M$. Our sample has a value of $e/a = 7.883$.

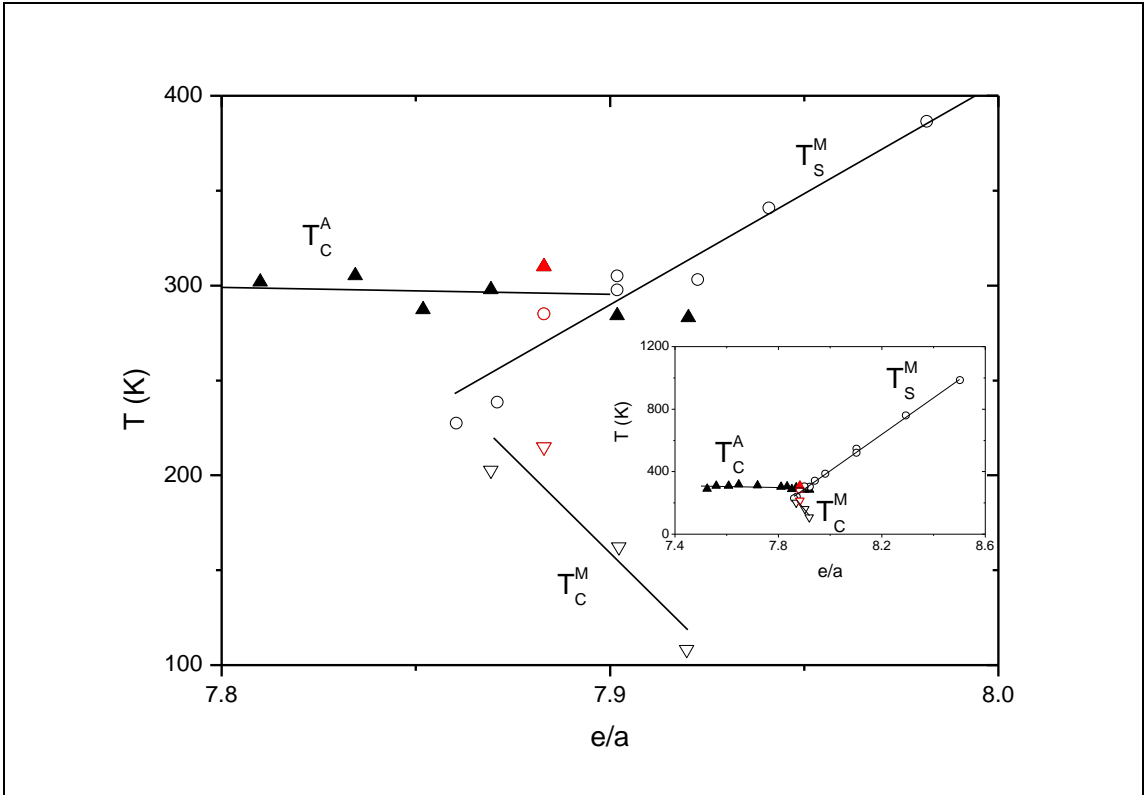

Figure 1: The magnetic and structural phase diagram of Ni–Mn–In Heusler alloys [4,15,16]. Triangles and circles correspond to magnetic and martensitic transformation temperatures, respectively. Data from this study have been added (red symbols). Inset shows the same phase diagram in a broader range of $e/a$.

3.1 MAGNETIC PROPERTIES

Figure 2a shows the specific magnetization $\sigma(H)$ as a function of the applied field *H* for some selected temperatures. At 100 K, ferromagnetic behavior is found for the martensite, which Curie temperature is estimated at $T_C^M$~215 K. Above this temperature, $\sigma(H)$ shows an almost linear behavior characteristic of the paramagnetic state of the martensite. Although deviations from the linearity are already observed for $\sigma(H)$ at 280 K for $\mu_0 H > 0.7$ T, magnetization abruptly increases at 285 K where $\sigma(H)$ shows a sharp rise at $\mu_0 H = 0.3$ T. As the magnetization experiments were performed on a heating protocol, the austenite start and finish temperatures could be obtained at 25 mT as $A_{start} = 282.2(2.5)$K and $A_{finish} = 292.2(2.5)$K, estimated as the temperature at which the magnetization rises and start to decrease, respectively. Increasing the magnetic field shifts these temperatures to lower values ($A_{start} = 277.1(2.5)$K and $A_{finish} = 284.8(2.5)$K at 1.5 T).

Both effects are ascribed to a field induced structural transition from paramagnetic martensite to ferromagnetic austenite. At this temperature, the magnetization shows steps that suggest that the ferroelastic transition does not occur smoothly but displays, instead, some jerky character. Above 285 K, magnetization decreases with temperature and, finally, a paramagnetic austenite is found above $T_C^A$ =310 K. Figure 2b shows the magnetization as a function of the temperature extracted from the $\sigma(H)$ measurements for two different fields. At low fields, the demagnetizing field yields an almost constant magnetization in the ferromagnetic ranges (from 50 to 190 K and from 290 to 305 K). In fact, at temperatures well below the Curie transition and low enough fields, the inverse of the magnetic susceptibility, $\chi_m$, can be negligible with respect to the demagnetizing factor, $N_D$ (with values between 0 and 1), and thus the specific magnetization $\sigma = \frac{1}{\rho}\chi_{app}H$ (where $\rho$ is the density and $\chi_{app} = \chi_m + 1/N_D$ is the apparent susceptibility) can be approximated by $\sigma \sim \frac{1}{\rho N_D}H$, which is constant for a constant field. At high fields $1/\chi_m \gg N_D$ (see Fig. 2a), being $\chi_{app} \sim \chi_m$ and the magnetization curve shows the expected thermal behavior in both ferromagnetic ranges. That is, $\sigma$ decreases as the temperature increases. However, the transitions occur in a broadened temperature range.

The values are within the reported ones for similar $e/a$. However, structural transition temperature is very sensitive in this $e/a$ range to small composition changes as can be observed in Figure 1.

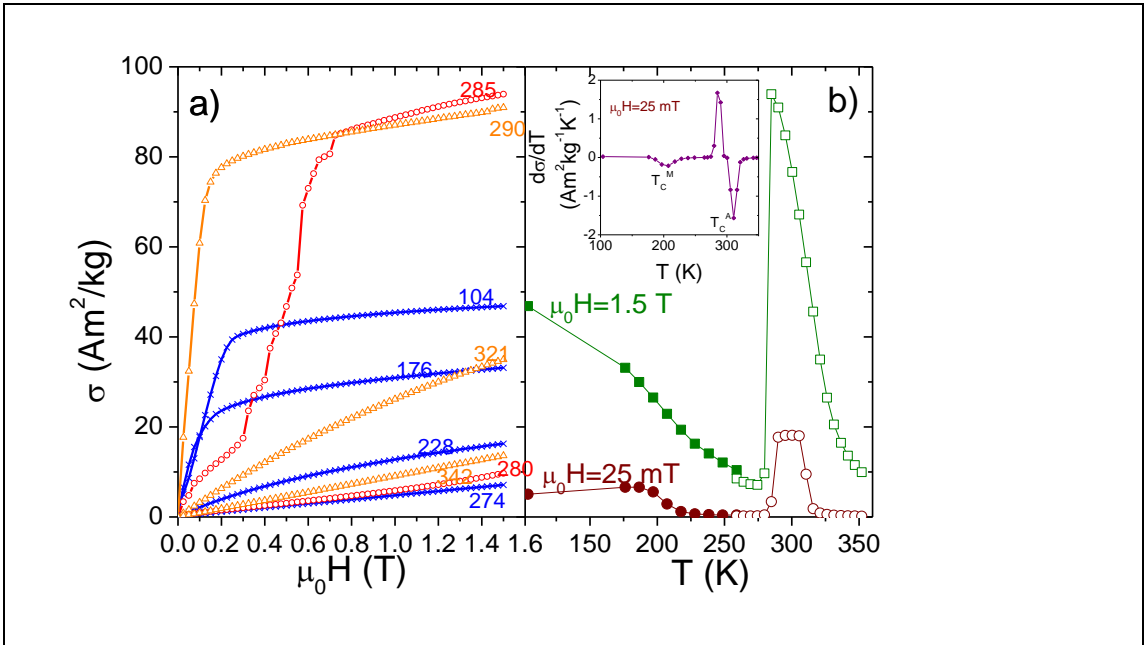

Figure 2. a) specific magnetization at different temperatures. Blue curves (crosses) correspond to the martensite phase; red curves (circles) correspond to two-phase behavior; and orange curves (triangles) correspond to austenite phase. b) temperature dependence of the specific magnetization at 25 mT and 1.5 T, respectively. Hollow symbols correspond to data obtained after zero cooling down to 150 K; the solid symbols were obtained at zero field cooling from the immediately above temperature (solid data point at 260 K, cooling from 352 K).

## 3.2 HEAT FLUX

Figure 3 shows DTA traces, defined as the ratio $\phi/r$ of the heat flux $\phi$ and temperature scanning rate $r$, for cooling and heating runs at $r=0.18$ K/h. $\phi/r$ gives the amount of released or absorbed energy per unit temperature. Two anomalies were detected, the first one at 310.5 K, the second one at 285 K. The first anomaly showed no hysteresis and looked like a small kink in the trace (see the insets in Figure 3). On the contrary, the second anomaly exhibited a broad extension with strong spikes that confirmed the jerky character of this transformation. For cooling experiments the anomaly extended from 291 K and 278 K, which correspond to the austenite start and end temperatures, while in heating runs the anomaly was located from 280 K to 294 K, which corresponds which correspond to the martensite start and end temperatures. The hysteresis in the temperature of the DTA trace maximum is 2.5 K, a signature of the first-order nature of this transition.

The temperatures of these anomalies agree with those determined by the magnetization measurements, which allows to associate both anomalies respectively with the ferromagnetic-paramagnetic transition of the austenite phase (high temperature) and the martensitic phase transition (low temperature).

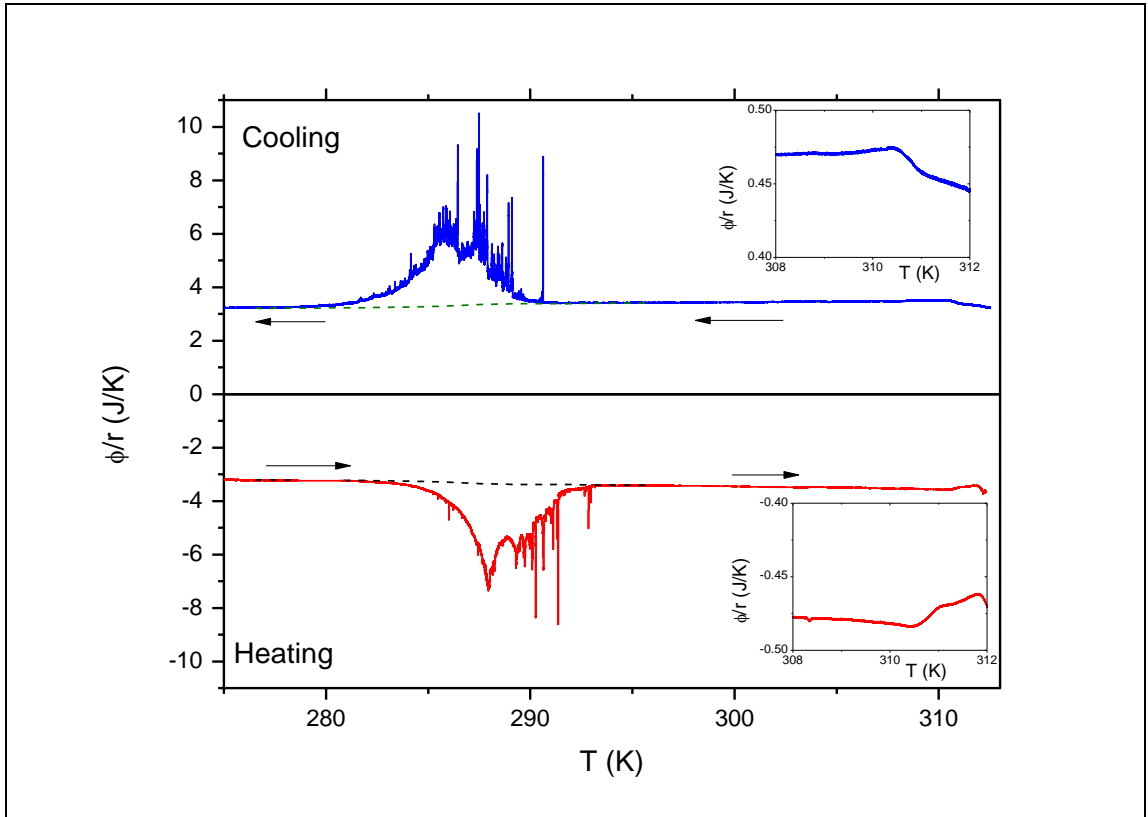

Figure 3. DTA traces as a function of the temperature. Dashed lines represent the baseline used to calculate the latent heat of the martensitic transition. The horizontal arrows indicate the direction of temperature change. The insets enlarge the small anomaly around 310.5 K.

3.3 SPECIFIC HEAT

The first-order character of the martensitic transition jeopardizes the experimental determination of the specific heat because, due to the latent heat, after a certain temperature is reached the conditions upon which the output signal of the calorimeter is linked to the specific heat of the sample are not fulfilled. The excitation energy is not stored kinetically as a sensible heat but instead it is also stored structurally as a latent heat, to which the specific heat is unrelated.

To solve this inconvenience, we developed a protocol for determining the specific heat of the sample during a heating run (r=0.04 K/h) taking advantage of the thermal hysteresis of this phase transition (2.5 K, as observed in the DTA traces). We applied a series of heat pulses with one in every four pulses being exceedingly large. Figure 4 shows a schematic representation of the expected evolution of the austenite (high temperature phase) molar fraction $x$ during the protocol.

The initial excitation increases the temperature of the sample from point 1 ($T_1$) to point 2 ($T_2$) and promotes the transformation into the austenite phase. When the heat power is cut off, the temperature of the sample relaxes to $T_3$ (point 3), the molar fraction is kept constant since the temperature $T_3$ is higher than the temperature at which the reversal phase transition takes place. Starting from temperature $T_3$, we measure the specific heat by applying several small amplitude thermal pulses (the temperature increment is around 1/20 smaller than the corresponding one for the first heat pulse) so that the temperature $T^*$ reached during these pulses does not overtake $T_2$ (see inset in Figure 4). As a result, the specific heat for a given phase composition of the sample can be determined. Then, a new large amplitude heat pulse promotes the temperature of the sample over $T_2$ until $T_4$ and the procedure is repeated at the new austenite phase fraction (see point 5 in Figure 4).

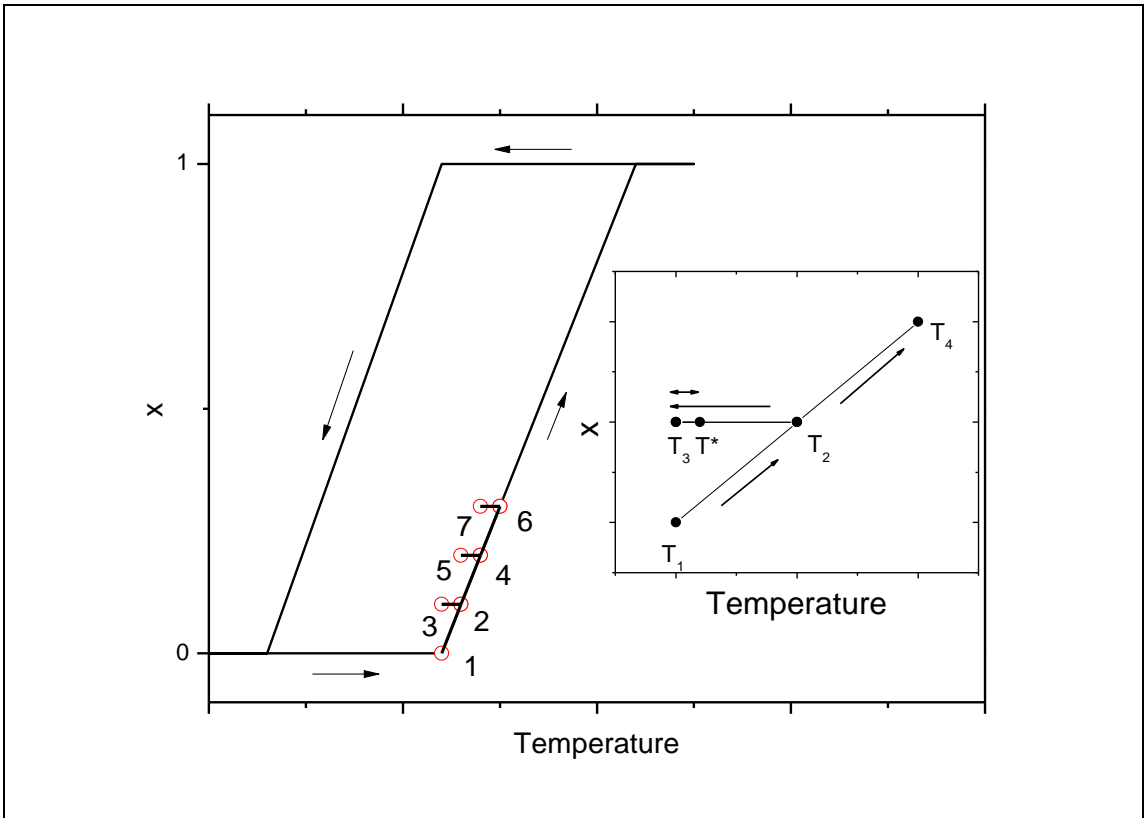

Figure 4. Schematic picture of the thermal hysteresis in the martensite-austenite phase transition (austenite phase fraction x versus temperature) during the measurement protocol for the specific heat. On a heating run the martensite-austenite transition is reached at point 1. A large thermal pulse heats the sample and promotes phase transformation until point 2. The pulse is switched off and the sample cools while thermal hysteresis freezes the reversal transformation. Point 3 is reached. There, smaller pulses can heat the sample back to T* (see the inset) and the specific heat for the given temperature can be determined. Then a large thermal pulse promotes back the phase transformation until point 4 is reached. The sample cools back to point 5 and the specific heat at the new temperature can be measured. The protocol continues until the phase transformation is completed.

Specific heat ($c_p$) data are shown in Figure 5. A lambda-type anomaly is observed around 310.5 K associated with the Curie point of the ferromagnetic-paramagnetic transition of the austenite phase. This anomaly is similar to that observed for $Ni_{45.37}Mn_{40.91}In_{13.72}$ [17,18] and $Ni_{49.3}Mn_{35.2}In_{15.6}$ [19] at similar temperatures. It is interesting to note that $T_C^A$ is weakly dependent on e/a. Around 287 K a step-like behavior is observed, associated with the martensitic phase transition. We will analyze separately the transitions in sections 3.4 and 3.6 while section 3.5 is devoted to the analysis of the avalanche criticalities.

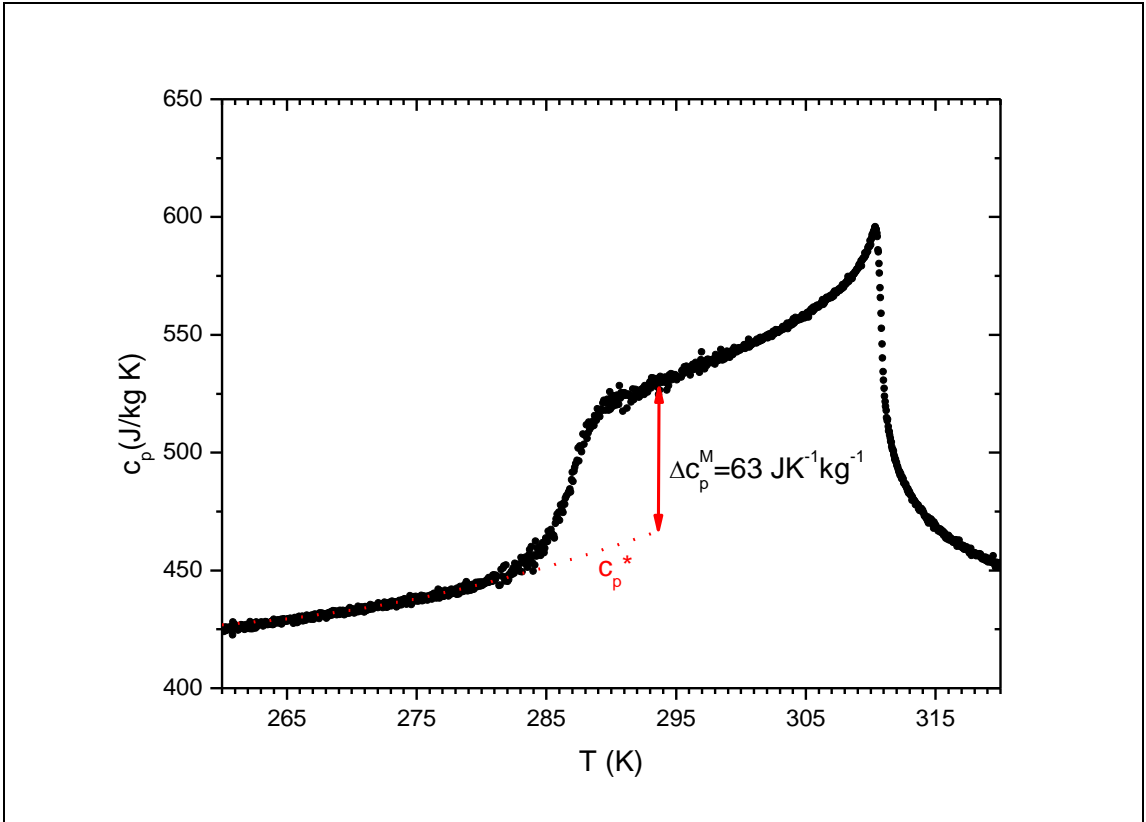

Figure 5. Temperature dependence of the specific heat for Ni$_{50.53}$Mn$_{33.65}$In$_{15.82}$ in the range across the ferroelastic phase transition around 287 K and the para-ferromagnetic phase transitions around 310.5 K. The dotted line was obtained by extrapolating the specific heat of the martensite phase for $T < 280$ K and was used to determine the jump in the specific heat associated with the ferroelastic transformation.

3.4 THE MARTENSITIC PHASE TRANSITION

The protocol used to measure c$_p$ is able to remove the impact of the latent heat and the intermittent dynamics during the phase transformation (see section 3.3). For $T < 280$ K the specific heat smoothly changes with the temperature. When the temperature increases and reaches the austenite start temperature, a smooth jump in the specific heat is observed from the value of $c_p = 445$ JK$^{-1}$kg$^{-1}$ at $A_{start} = 280$ K to the value $c_p = 524$ JK$^{-1}$kg$^{-1}$ at $A_{finish} = 294$ K.

A similar jump in c$_p$ has been observed for the martensitic phase transitions in Ni$_{45.37}$Mn$_{40.91}$In$_{13.72}$ [17,18] and Ni$_{49.3}$Mn$_{35.2}$In$_{15.6}$ [19]. In those works the step-like behavior is accompanied by a peak, which is not observed in our measurements. We conjecture that this peak is related to the latent heat released during the transformation and which is removed by our measurement protocol.

The change of the specific heat associated to the onset of the ferromagnetic austenite phase can be determined from our data by computing the difference between the experimental data and the value given by an extrapolation of the specific heat for T<280 ($c_p^*$ in Figure 5). At $A_{finish}$ this yields $\Delta c_p^M = 63$ JK$^{-1}$kg$^{-1}$ Integrating the excess specific heat between $A_{start}$ and $A_{finish}$, the enthalpy excess associated to specific heat variation was determined as $\Delta H_c = 216(1)$ Jkg$^{-1}$. By scaling it with $T_0 \cong 287.5$ K the excess entropy was $\Delta S_c \cong \Delta H_c/T_0 = 0.751(4)$ JK$^{-1}$kg$^{-1}$is obtained. This increase of entropy may be ascribed to the activation of new modes related to the interaction between magnetic moments in the ferromagnetic phase.

Nevertheless, the total enthalpy excess of the transition must include the contribution of the latent heat, which is computed by integrating the DTA trace respect to an appropriate baseline (dashed lines in Figure 3) calculated from specific heat data [20]. The computed transition enthalpy is represented in Figure 6 for the heating and cooling runs. The average value is $\Delta H_t = 7.21(15)$ kJkg$^{-1}$, corresponding to heating and cooling runs. Notice that the enthalpy attributed to the latent heat is 30 times larger than the excess enthalpy attributed to the excess specific heat. The entropy excess is estimated as $\Delta S_t \cong \Delta H_t/T_0$, where $T_0$, defined as the average temperature between the heating and cooling loops, is 288 K, which yields $\Delta S_t = 25.0(5)$ J$K^{-1}$kg$^{-1}$. These values are in good agreement with reported data for other Ni-Mn-In compositions. [4,21]

Figure 7 shows the dependence of the entropy excess with the electronic concentration *e/a*. It has been plotted with data taken from the literature [4,21]. In region I of the phase diagram the martensitic transition takes place between paramagnetic phases; $\Delta S$ has a vibrational origin and is practically independent of *e/a*. In region II the martensitic transition takes place between a high temperature ferromagnetic phase and a low temperature paramagnetic phase. Consequently, $\Delta S$ has a magnetic contribution of opposite sign to the vibrational one. This contribution increases as the martensitic transition moves away from the Curie temperature, which explains the decrease in $\Delta S$ when *e/a* decreases.

3.5 AVALANCHE CRITICALITY

The large number of spikes (Figure 3) is a signature of the DTA traces in the martensitic transformation of some shape memory alloys. As an example a similar behavior was found in some Cu-Al-X shape memory alloys[22–25]. Martensitic transitions happen through a succession of metastable states[26] and display avalanche criticality[27]. Although avalanches are typically detected by acoustic emission measurements[8], they have also been detected by high precision calorimetry[22–25]. The spikes observed in the DTA traces represent quasi-instantaneous heat flux peaks which characterize at a macroscopic scale the energies associated with the intermittent dynamic of the phase transition and the latent heat during the relaxation between metastable states.

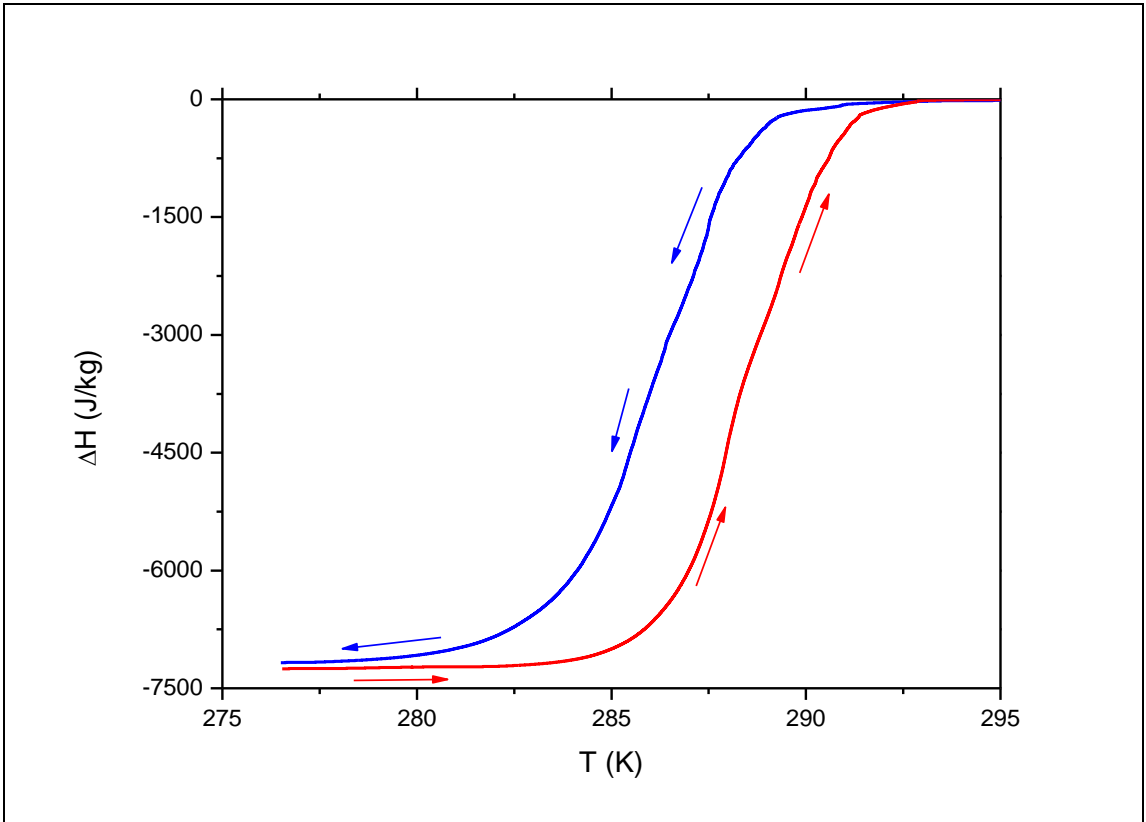

Figure 6. Excess enthalpy computed after integration of the normalized heat flux shown in Fig. 3. The thermal hysteresis ($\Delta T \sim 2.5$ K) is accompanied by different rounding mechanisms.

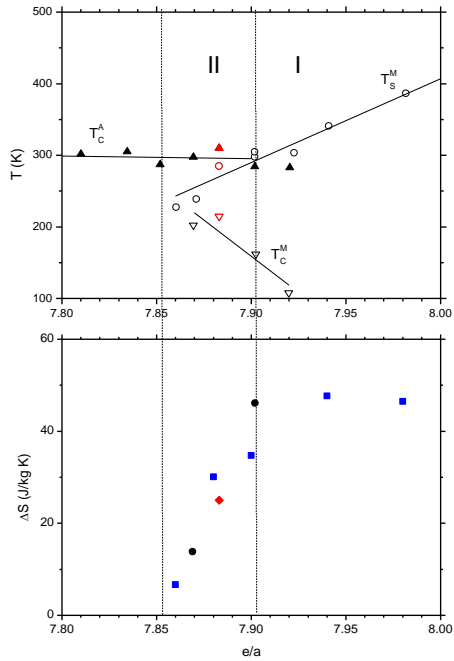

Figure 7. Phase diagram (top) and dependence of the entropy excess $\Delta S$ (bottom) [4,21] with the electronic concentration $e/a$. Data from this study have been added (red symbols). In region I the martensitic transition takes place between paramagnetic phases and in region II the martensitic transition takes place between a high temperature ferromagnetic phase and a low temperature paramagnetic phase.

The spikes of the DTA signal can be statistically characterized by grouping events by size. For this purpose, a fifth-order all-pole Butterworth filter with a normalized cut-off frequency of $8 \times 10^{-2}$ Hz smoothed the signal. Then we searched for local maxima and local minima. The size of each event was determined by the monotonous increase of the signal from a local minimum to the following local maximum. Introducing the sensibility and the time constant of the calorimeter ($\tau = 78$ s) the event can be energetically characterized.

Event sizes were then grouped in logarithmic bins. Figure 8 shows the experimental estimation of the probability density of calorimetric events larger than $E_0 = 17$ µJ versus event size for the cooling (upper panel) and heating (lower panel) experiments. Data can be fitted to a power law $P(E) = aE^{-\varepsilon}$ with avalanche exponents $\varepsilon = 2.03(6)$ (cooling) and $\varepsilon = 2.04(4)$ (heating).

To confirm the avalanche behavior obtained from calorimetric measurements, we have performed AE measurements, which is the usual experimental technique used to characterize avalanche criticality in ferroelastic transitions. From the results at selected cooling and heating rates, plots showing fitted avalanche exponent ε by means of a maximum likelihood analysis as a function of a varying lower cut-off $E_{min}$ are obtained. A power law behavior should lead to a significant plateau that defines the corresponding critical avalanche exponent. [28] The analysis is presented in Figure 9 and confirms a good power law behavior with an avalanche exponent ε = 2.00 ± 0.15 for both heating and cooling runs, which is independent of the temperature rate in the range from $10^{-2}$ K/s to $10^{-1}$ K/s. Within the errors, the obtained AE exponent compares well with the one reported for Ni-Mn-Ga in Refs. [10,29] from AE measurements in the absence of applied stress. It is also consistent with the value determined from calorimetric measurements. Taking into account that both AE and calorimetric techniques are able to detect avalanches at very different energy ranges, this consistency corroborates scale invariance. Moreover, the exponent obtained is in very good agreement with the expected critical avalanche exponent, $\varepsilon$ = 2, for systems that transform from a high temperature cubic phase to low temperature monoclinic phase, for which a variant multiplicity of 12 may occur.

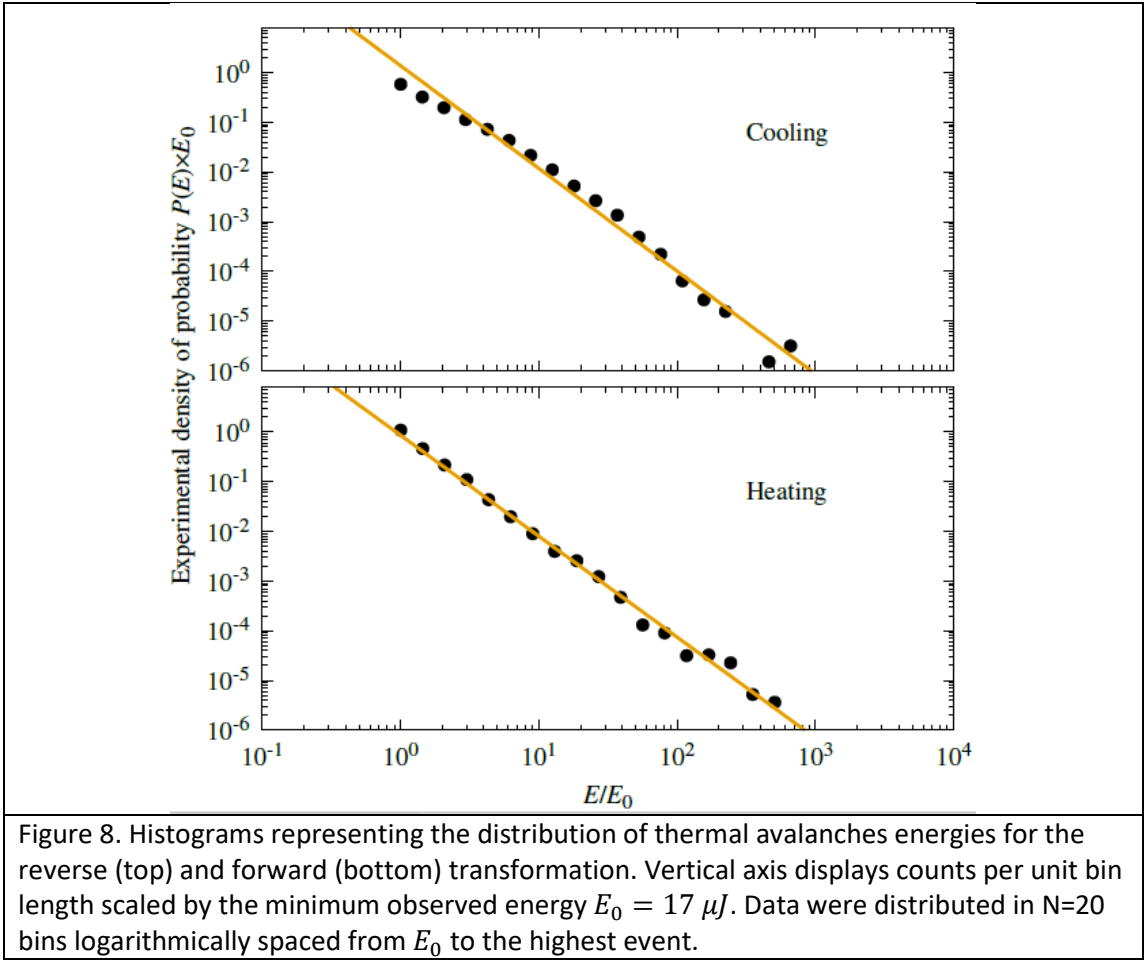

Figure 8. Histograms representing the distribution of thermal avalanches energies for the reverse (top) and forward (bottom) transformation. Vertical axis displays counts per unit bin length scaled by the minimum observed energy $E_0 = 17 \ \mu J$. Data were distributed in N=20 bins logarithmically spaced from $E_0$ to the highest event.

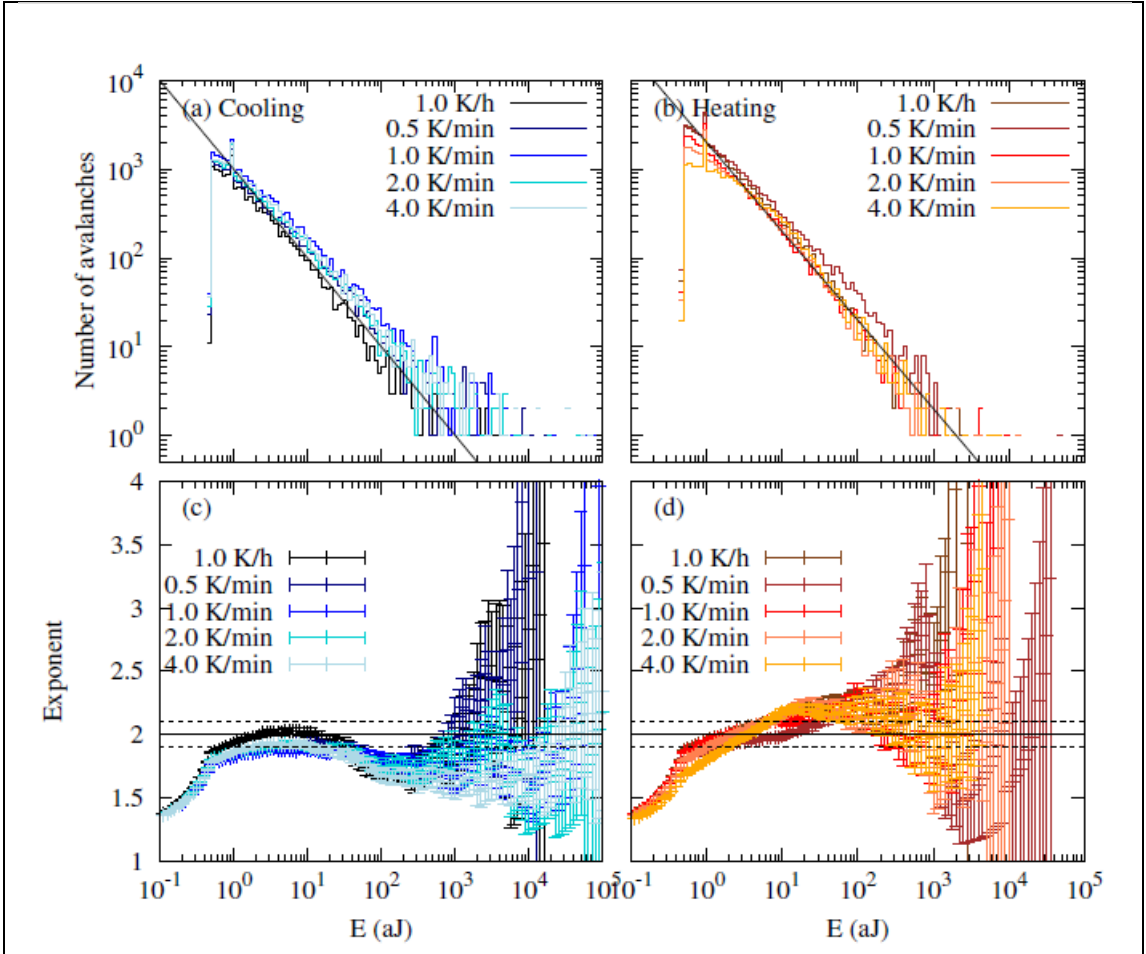

Figure 9. Fitted exponent as function of the energy $E$ for cooling and heating AE runs at selected temperature rates. The meaning of $E$ is in fact different in top and lower panels. In top panel it corresponds to the energy of the avalanches while in lower panel it represents energy cut off of the maximum likelihood fitting. In the lower panels, the horizontal lines give the value of the avalanche exponent (continuous line) and the corresponding error bar (dashed lines).

3.6 THE CRITICAL POINT

The ferromagnetic-paramagnetic transition for the austenite phase describes on the heat capacity an asymmetric lambda-type anomaly around 310.5K. The asymmetry is noted in Figure 5 by larger values of the specific heat in the ferromagnetic low-temperature phase compared to those of the paramagnetic high-temperature phase. This might suggest differences in the critical behavior for every phase.

With $t = (T - T_c)$ and following Kornblitt and Ahlers[30] and Marinelli[31] fitting method we tested the model:

$$c(T) = (A/\alpha)t^{-\alpha}(1 + D|t|^{0.5}) + B + Et \qquad [1]$$

where $B + Et$ is a regular contribution and $(A/\alpha)t^{-\alpha}(1 + D|t|^{0.5})$ is a critical contribution plus a confluent singular term. In what follows primed quantities refer to the $t < 0$ branch and non-primed values refer to the $t > 0$ branch.

We imposed the restrictions $B = B$, and $E = E'$ so that the regular contribution keeps still for every phase and $\alpha = \alpha'$ and $T_c = T_c'$ so that the critical divergence also keeps still for every phase. For $\alpha = 0.09$ experimental data and the tested model showed extremely good agreement for $|\Delta t|\sim 0.03 T_c$ (the mean residual is 9.2 JK$^{-1}$kg$^{-1}$). The parameters for Eq. 1, determined by the Marquadt-Levenberg algorithm available in Gnuplot, are listed in Table 1. Data in the colored strip-band centered around the temperature of the highest specific heat data and $2 \times 10^{-3} T_c$ in amplitude were excluded from the fit (see Figure 10). We also tested $\alpha \geq 0.1$ and $\alpha \leq 0.07$ but this resulted in physically inconsistent regular contributions. The blue line either crossed the experimental data above $T_c$ (if $\alpha \geq 0.1$) or crossed the ideal vibrational contribution [32] below $T_c$ (if $\alpha \leq 0.07$).

The confluent singular term impacts the fit in two ways. First it helps to account for the rounding of the peak. Notice that the model deviates from the experimental data at $T\sim T_c$ in the paramagnetic high-temperature phase and resumes its agreement with experimental data just when the maximum experimental value is observed. Secondly and noticing $D/D' \sim -10$ the confluent term heavily modifies the critical divergence $\alpha = 0.09$ in the paramagnetic high-temperature phase pulling it all the way down (notice $D < 0$) to the experimental data. Comparatively, the confluent term only slightly modifies the critical divergence in the ferromagnetic phase.

Previous observations of the specific heat of Ni compounds [33,34] yielded a negative critical exponent $\alpha$ as expected for a Heisenberg magnet. Our Heusler alloy seems to show a critical behavior ($\alpha = 0.09$) different to Ni but similar to Ni-Mn-Ga ($\alpha = 0.03$)[35]. In any case the fitted value of $(1 - A/A')/\alpha$ is 4.0 (see Table 1) in agreement with data reported for other systems including Ni [34]. To understand the differences in the critical behavior, the inset in Figure 10 shows the specific heat for Ni[33,35] (green) and for the alloys Ni-Mn-Ga[35] (vermillion) and Ni-Mn-In (yellow [18]; black, this work) and visualizes different critical divergence in the specific heat of Ni as compared to other Heusler alloys. This eventually leads to differences in $\alpha$.

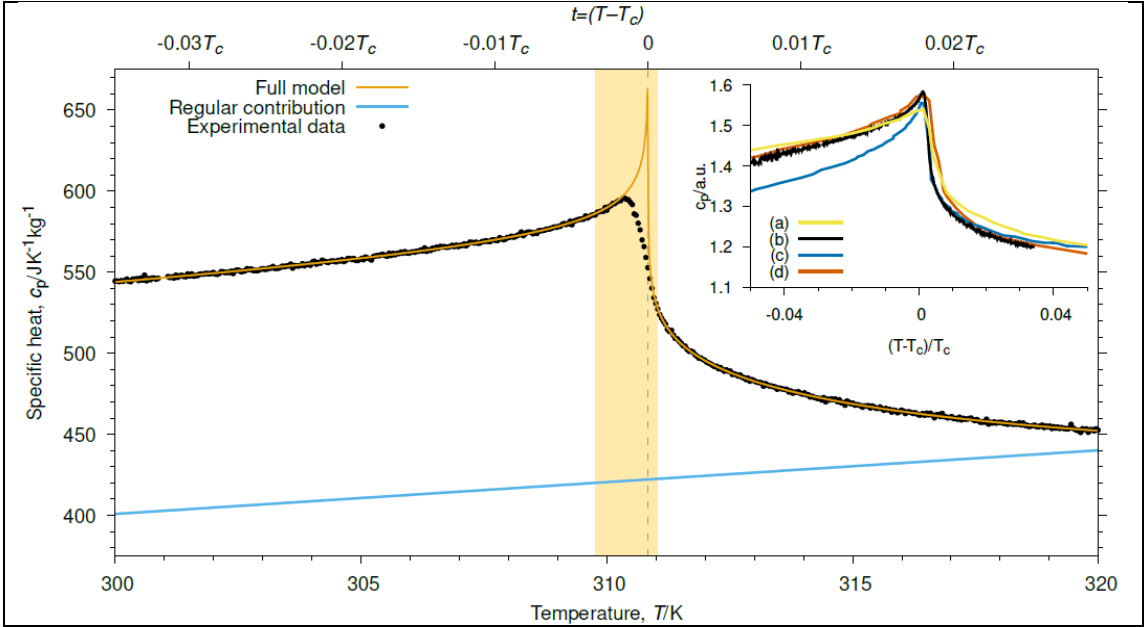

Figure 10. The specific heat of $Ni_{50.53}Mn_{33.65}In_{15.82}$ around the ferromagnetic-paramagnetic phase transition of the austenite phase (circles) and the model of Equation [1] (orange line) with a critical divergence ($\alpha = 0.09$ for both phases), a confluent singular term and a regular contribution (blue line). The vertical strip band signals the region excluded from the fitting. The vertical dashed line signals $T_c$, obtained from the fitting. The inset shows the specific heat of: a) Ni-Mn-In from Ref. [18], b) Ni-Mn-In from this work (black), c) Ni (blue) from Ref[33,35] and d) Ni-Mn-Ga (vermillion) from Ref. [35] versus reduced temperature $(T-T_c)/T_c$. The inset shows that the critical behavior of Heusler alloys differs from the Heisenberg 3D divergence shown by pure Ni[34].

**Table I**

| Parameter | Value |
| --- | --- |
| $\alpha = \alpha'$ | 0.09 (fixed) |
| A | 9.309 |
| A' | 14.625 |
| B=B' | 421.81 |
| E=E' | 1.960 |
| D | -0.283 |
| D' | 0.0275 |
| $T_c$ | 310.8439 |
| A/A' | 0.6365 |
| D/D' | -10.306 |
| $(1-A/A')/\alpha$ | 4.0 |

Table I: parameters obtained by least-square fitting of the data. The resulting units of $c_p$ are $JK^{-1}kg^{-1}$ when the unit of the temperatures is kelvin.

4. CONCLUSIONS

Magnetization measurements revealed that $Ni_{50.53}Mn_{33.65}In_{15.82}$ transforms on cooling from a paramagnetic austenite phase into a ferromagnetic austenite, then into a paramagnetic martensite and finally into a ferromagnetic martensite when the temperature is decreased. Specific heat and Differential Thermal Analysis characterized energetically the first two transformations: the magnetic transition for the austenite phase and the martensitic phase transition.

The difference between the specific heat in the ferromagnetic austenite phase and the paramagnetic martensite phase was characterized by a step-like anomaly. For this, we took advantage of the thermal hysteresis of this phase transition and the specific heat was determined avoiding any disturbance coming from structural energetic changes (latent heat). The entropy excess agreed with previous reported values and with the fact that the phase transition takes place in the region where the magnetic and vibrational contributions to entropy oppose to each other.

The intermittent nature of the martensitic transformation showed similar power law distributions for thermal and for acoustic emission measurements despite both techniques being sensitive to very different energy scales. The critical exponent was found close to previous values reported for Ni-Mn-Ga alloys. Likewise, the critical behavior of the specific heat anomaly at the paramagnetic-ferromagnetic austenite phase transition showed similarity with previous values reported for Ni-Mn-Ga alloys and markedly different from data previously reported for Heisenberg type systems such as pure Ni. Further research should inspect whether this difference is a characteristic of Heusler alloys.

ACKNOWLEDGEMENTS

This work was supported by the PAI of the Regional Government of Andalucía (grants to research groups FQM-130 and FQM-121). JB acknowledges support from AEI/FEDER-UE (Projects US-1260179 and P18-RT-746) and EV and AP from CICyT (Spain), Project No. MAT2016-75823-R.